\documentclass[letterpaper]{article} 
\usepackage{aaai2026}  
\usepackage{times}  
\usepackage{helvet}  
\usepackage{courier}  
\usepackage[hyphens]{url}  
\usepackage{graphicx} 
\usepackage{booktabs}
\usepackage{censor}
\usepackage{placeins}
\usepackage{xcolor}
\usepackage{natbib}  
\usepackage{caption} 
\usepackage{todonotes}
\usepackage{algorithm}
\usepackage{algorithmic}
\usepackage{tabularx}

\usepackage{xcolor}

\usepackage{subcaption}
\usepackage{newfloat}
\usepackage{listings}
\DeclareCaptionStyle{ruled}{labelfont=normalfont,labelsep=colon,strut=off} 
\floatstyle{ruled}
\newfloat{listing}{tb}{lst}{}
\floatname{listing}{Listing}
\title{WhatsApp Vaccine Discourse (WhaVax): An Expert-Annotated Dataset and Benchmark for Health Misinformation Detection*}
\author{
    Jônatas H. dos Santos\textsuperscript{\rm 1}, Julio C. S. Reis\textsuperscript{\rm 2}, Philipe F. Melo\textsuperscript{\rm 2}, João F. H. Olivetti\textsuperscript{\rm 2}, Thales H. Silva\textsuperscript{\rm 1}, Matheus Gontijo Guimaraes\textsuperscript{\rm 1}, Glaucio de Souza\textsuperscript{\rm 1}, Marcos A. Gonçalves\textsuperscript{\rm 1}, Fabricio Benevenuto\textsuperscript{\rm 1}, Filipe B. B. Zanovello\textsuperscript{\rm 1}, Marco A. G. Rodrigues\textsuperscript{\rm 1}, Cristiano X. Lima\textsuperscript{\rm 1}\\
}
\affiliations{
    \textsuperscript{\rm 1}Universidade Federal de Minas Gerais (UFMG), Brasil\\ 
    \textsuperscript{\rm 2}Universidade Federal de Viçosa (UFV), Brasil\\

    jonatashds@ufmg.br, jreis@ufv.br, philipe.freitas@ufv.br, joao.olivetti@ufv.br, thaleshenrique@ufmg.br\\
    guimaraesmatheus22@gmail.com, glaucio.gss@gmail.com,
    mgoncalv@dcc.ufmg.br, fabricio@dcc.ufmg.br, filipebessa9@gmail.com, magro.mg@terra.com.br, cxlima@ufmg.br
}

\begin{document}

\maketitle

\begin{abstract}
We introduce \textbf{WhaVax}, a new expert-annotated dataset of vaccine-related WhatsApp messages collected from large Brazilian public groups spanning multiple pandemic years. The dataset was constructed through a rigorous, carefully designed pipeline that integrates keyword-based data collection, semantic deduplication to remove near-duplicate content, and a multi-stage annotation protocol conducted by medical specialists. This process produced a high-quality gold-standard corpus, characterized by substantial inter-annotator agreement and strong reliability for downstream analysis. Additionally, we provide a detailed characterization of WhatsApp misinformation, revealing distinctive linguistic, structural, lexical, temporal, and group-level patterns, as well as a meaningful layer of ambiguous cases that reflect the complexity of health discourse in private messaging. We also benchmark classical models, fine-tuned Small Language Models, and zero- or few-shot Large Language Models under realistic data-scarcity constraints, demonstrating that strong embeddings and LLM approaches perform competitively, while domain alignment and data availability remain critical factors. This study provides a rare, high-quality resource to support misinformation research and computational modeling in encrypted communication environments.\looseness=-1
\end{abstract}

\section{Introduction} \label{sec:introduction}

Brazil currently faces a renewed public health alert: the country has returned to the list of the twenty nations with the highest number of unvaccinated children worldwide, according to UNICEF and WHO~\cite{UNICEF_WHO_2025}. This decline poses a serious risk for a nation historically recognized for successful mass immunization campaigns, and health misinformation has emerged as a key factor undermining confidence in vaccines and public health institutions \cite{10.1093/jamia/ocab219}.\looseness=-1

Within this ecosystem, WhatsApp plays a central role~\cite{benevenuto2024misinformation}. With more than 120 million users in Brazil and deep penetration across regions and demographics, the platform’s encrypted and semi-private group structure enables misinformation \cite{sbbd_vaccine, reis2023helping} to circulate rapidly while limiting monitoring and moderation capabilities \cite{melo2019complexnet,reis2020can}. During the COVID-19 pandemic, WhatsApp became a particularly active arena for rumors, conspiratorial narratives, and alarmist vaccine discourse, amplifying uncertainty and distrust \cite{Sharma2023-dr}.\looseness=-1

Despite its undeniable societal impact, misinformation research on WhatsApp faces unique structural and technical barriers.
Access restrictions, privacy concerns, informal language, and the highly contextual nature of messaging make systematic analysis exceptionally difficult~\cite{garimella2025globalpatternsviralcontent}. As a result, the research community still lacks high-quality, expert-validated datasets that capture how health misinformation actually unfolds in WhatsApp conversations. Without such resources, it is difficult to characterize misinformation dynamics, evaluate computational approaches, or inform public health interventions.\looseness=-1

This paper addresses this gap by constructing and analyzing a new expert-annotated dataset of vaccine-related WhatsApp messages. Developed through a rigorous multi-stage medical annotation process, the dataset spans multiple years of pandemic discourse and reflects misinformation as it appears in authentic conversational environments, rather than in curated fact-checking portals or public social platforms. Building this dataset raises broader research questions (RQs) that motivate our study:\looseness=-1

\begin{itemize}
\item \textbf{RQ1}: Is it feasible to construct a domain-specific, expert-annotated, high-quality dataset of medical misinformation based on authentic WhatsApp communication?
\item \textbf{RQ2}: Given the inherent constraints of such environments -- limited samples, informal text, and topic specificity -- what limitations and opportunities arise when training classifiers on small, specialized datasets?
\item \textbf{RQ3}: How do different classification strategies (classical machine learning, fine-tuned Transformer-based Small Language Models (SLMs), and zero and few-shot Large Language Models (LLMs)) compare when applied to a medical misinformation classification task in WhatsApp?
\end{itemize}

To address these questions, we (i) build and publicly release an expert-labeled dataset of WhatsApp vaccine-related messages, called \textbf{WhaVax}; (ii) provide a detailed characterization of linguistic, structural, lexical, temporal, and group-level patterns; and (iii) systematically compare classical classification models, Small Language Models (SLMs), and Large Language Models (LLMs) under realistic data-scarcity conditions for misinformation classification.\looseness=-1

Our results show that constructing such a dataset is feasible and highly informative. Careful filtering, semantic deduplication, and expert review yield a reliable gold-standard corpus with meaningful agreement. Despite size and access constraints, the dataset enables rich analyses, revealing distinct stylistic and behavioral patterns, as well as meaningful ambiguity zones that reflect the real complexity of health misinformation. Furthermore, our modeling experiments demonstrate that even small, carefully curated datasets can support effective automatic detection. Classical models with strong embeddings remain competitive, fine-tuned Transformers are highly sensitive to domain alignment, and zero-/ few-shot LLMs achieve strong performance without supervised tuning, although outcomes vary across architectures.\looseness=-1

In sum, this work offers three main contributions: (1) a novel, expert-annotated dataset of vaccine-related WhatsApp messages; (2) an in-depth empirical characterization of misinformation in group messaging environments; and (3) a systematic comparative evaluation of misinformation classifiers under realistic, resource-constrained conditions. Together, these contributions provide a rare, high-quality resource for the community and actionable evidence to support computational research and public health strategies in encrypted messaging platforms.\looseness=-1

\section{Related Work and Datasets} \label{sec:related}
The study of online health information has received increasing attention, particularly during the COVID-19 pandemic \cite{covid_misinfo_1}, which amplified previously niche anti-vaccination narratives \cite{antivax_desinfo_1, antivax_desinfo_2}. These developments highlighted the societal risks of misinformation, especially in crisis contexts \cite{covid_misinfo_2}, where misleading content directly conflicts with medical guidance and can endanger lives.\looseness=-1

\vspace{0.1cm}
\noindent\textbf{Misinformation on Online Social Networks.}
The ubiquity of online information has reshaped everyday life, influencing behaviors \cite{internet_behavior} and democratic processes worldwide \cite{elections1, elections2, elections3}. However, the volume and speed of information circulation also create conditions that complicate truth-seeking. Research on information overload shows that users exposed to excessive content are more likely to share unverified or incorrect information \cite{overload1, overload2}. Social media platforms, in particular, have been repeatedly identified as major vectors of online misinformation \cite{osn_misinfo1, osn_misinfo2}.\looseness=-1

\vspace{0.1cm}
\noindent \textbf{Datasets on Vaccine Opinion and Misinformation.}
A systematic review on Instant Messaging Misinformation highlights automatic detection as a key research direction on the field \cite{webmedia_disinfo_review}. However, this remains challenging due to the open-ended nature of misinformation, which varies in topic, format, and degree of subtlety. Progress in this area depends critically on access to diverse, high-quality datasets that capture both reliable and misleading information across platforms. To address this need, several studies have released publicly available datasets on vaccine discourse and credibility assessment. Table \ref{tab:vaccine_misinfo_datasets} summarizes these resources, highlighting their scope, content, and annotation strategies.\looseness=-1

\begin{table*}[ht]
\centering
\scriptsize
\begin{tabularx}{\textwidth}{p{2cm}p{6cm}p{3.5cm}p{4.5cm}}
\hline
\textbf{Dataset} & \textbf{Description} & \textbf{Labels} & \textbf{Annotation Method} \\
\hline
     \cite{cui2020coaid} &  COVID-19 healthcare misinformation dataset with news articles, social media posts, and user engagements. &  Fake vs. real + metadata labels. &  Fact-check verification. \\\hline
     \cite{hossain-etal-2020-covidlies} &  ~6.6k tweets annotated for stance regarding curated COVID-19 misconceptions. &  Agree/Disagree/No Stance per misconception. &  Misconceptions curated by medical experts + human annotation. \\ \hline
     \cite{weinzierl-covid-glp} &  Tweets mapped to specific COVID-19 vaccine misinformation targets; includes stance labels. &  Target labels; stance: Agree/Disagree/No Stance. &  Curated misinformation targets + human expert annotation. \\\hline
     \cite{chen2021mmcovar} &  Multimodal repository with news and tweets (text + images + temporal data) focused on COVID-19 vaccines. &  News: reliable/unreliable; Tweets: reliable/unreliable/inconclusive. &  Media credibility rankings + manual stance/credibility annotation. \\\hline
     \cite{Hayawi2022ANTiVax} &  Large Twitter dataset (\>15M tweets) with ~15k manually annotated for COVID-19 vaccine misinformation. &  Binary: misinformation vs. general vaccine tweet. &  Human annotation + expert validation. \\\hline
     \cite{Crupi_Mejova_Tizzani_Paolotti_Panisson_2022} &  Twitter dataset (16.2M tweets) of the Italian COVID-19 vaccination debate (665k unique users, Sep 2019--Nov 2021). &  Stance: Supporter/Hesitant/Other/Pets. &  Manual annotation propagated via hierarchical clustering on retweet networks.\\\hline
     \cite{Poddar_Mondal_Misra_Ganguly_Ghosh_2022} &  Large Twitter dataset (15.7M tweets) for long-term (Jan 2018–Mar 2021) analysis of vaccine opinions, focusing on stance change over time. &  Tweet Stance: Anti-Vax/Pro-Vax/Neutral (used for user classification). &  Human annotation (1.7k tweets) + CT-BERT++ classifier for user stance categorization (97\% precision). \\\hline
     \cite{Weinzierl_Harabagiu_2022} &  Large Twitter dataset (9.1M original tweets, Dec 2019-Jul 2021) used to identify 113 Vaccine Hesitancy Framings and derive 9 user profiles. &  Framing Stance: Accept/Reject/Doubt. 9 User Hesitancy Profiles. &  Tweets are summarized through a QA system; Hesitancy profiles are clustered through sparse k-means.\\\hline
     \cite{Giovanni_Pierri_Torres-Lugo_Brambilla_2022} &  Large Twitter dataset (over 70M tweets) on COVID-19 vaccine conversations in French, German, and Italian (Nov 2020–Nov 2021).&  Stance: Pro-vaccines/Anti-Vaccines/Neutral/Out-of-Context. &   Annotation by native speakers of 1000 random tweets per language; a third annotator resolved conflicts. \\\hline
     \cite{poddar2022caves} &  Large-scale dataset (10k tweets) for multi-label classification and summarization of COVID-19 anti-vaccine concerns. &  12 specific concern classes (multi-label) + explanations for each label. &  Manual annotation by human experts (labels and natural language explanations). \\ \hline
     \cite{zarei2023vaxculturedatasetstudyingvaccine} &  6,373 vaccine-related tweets annotated with stance, misinformation indicator, entities, and message type. &  Stance (pro/anti/neutral), misinformation (yes/no), entities, message type. &  Manual annotation by trained communicators/journalists. \\\hline
     \cite{Mu_Jin_Grimshaw_Scarton_Bontcheva_Song_2023} &  Dataset of 3,101 tweets in English on attitudes toward COVID-19 vaccination, notable for separating hesitancy from anti-vaccine attitude. &  Pro, Anti, Hesitancy, Irrelevant. &  Manually annotated into four categories; the process included annotator training and quality testing. \\\hline
\end{tabularx}
\caption{Datasets related to vaccine misinformation.}
\label{tab:vaccine_misinfo_datasets}
\end{table*}

\vspace{0.1cm}
\noindent \textbf{Research Gap.}
As summarized in Table~\ref{tab:vaccine_misinfo_datasets}, most publicly available vaccine misinformation datasets are sourced from Twitter/X. While these resources are valuable, they provide a skewed perspective focused exclusively on public-facing discourse on Twitter/X. As a result, they fail to capture the unique dynamics of group communication on WhatsApp, which have become central to modern misinformation ecosystems. Unlike public social platforms, WhatsApp feature end-to-end encryption and a semi-private group structure that enables the rapid spread of rumors while hindering external monitoring or moderation. Furthermore, by introducing a large-scale corpus in Portuguese, \textbf{WhaVax} bridges the gap between public and private platform analysis and enhances linguistic diversity, enabling more robust cross-cultural and cross-lingual investigations into vaccine discourse.

\section{Dataset Construction Methodology} \label{sec:metho}


Our dataset is based on the publicly available WhatsApp repository introduced in \cite{Resende-WWW2019, melo2019whatsapp}, which compiles messages from large Brazilian public groups. The content covers 2020–2023 and comes from public groups accessible through invitation links obtained via community-sharing platforms and targeted Web searches, initially guided by Brazilian political keywords.

Data collection was conducted using an automated large-scale pipeline, where multiple registered accounts continuously retrieved group messages. For each message, both text and metadata were collected, including unique identifiers, sender information, group origin, media type, and timestamps. In accordance with standard practices in instant messaging research, all personally identifiable information was anonymized to comply with Brazilian data protection regulations, retaining only area codes to allow coarse-grained geographic analysis without compromising privacy. Processed messages were stored in JSON format to support subsequent analysis.\looseness=-1

\subsection{Health Data Filtering}

The original repository contains millions of messages collected over three years, most of which are unrelated to vaccination or health. To isolate relevant content, we applied a keyword-based filtering procedure using Portuguese vaccination terms (e.g., ``vacina'', ``vacinação'', ``vacinado'') along with variant spellings, slang, and obfuscated forms (e.g., ``v4c1n4'', ``vachina''), which are commonly linked to conspiratorial discourse. Additional terms, including vaccine brand names, were tested but later discarded after yielding a high proportion of off-topic content.\looseness=-1

\begin{table}[t]
\centering
\scriptsize
\vspace{-0.3cm}
\resizebox{0.9\linewidth}{!}{%
\begin{tabular}{ l r r r }
\toprule
\vspace{-0.05cm}
\textbf{keyword} & \textbf{\#Occurrences} & \textbf{Unique users} & \textbf{Unique groups} \\
\midrule
\vspace{-0.1cm}  vacina      &  65373 &  10710 &  876 \\
\vspace{-0.1cm}  vacinas     &  32476 &  6835  &  768 \\
\vspace{-0.1cm}  vacinação   &  17773 &  4450  &  695 \\
\vspace{-0.1cm}  vacinados   &  7813  &  2521  &  533 \\
\vspace{-0.1cm}  vacinar     &  5675  &  2458  &  532 \\
\vspace{-0.1cm}  vacinado    &  3504  &  1758  &  473 \\
\vspace{-0.1cm}  vacinadas   &  2517  &  1058  &  365 \\
\vspace{-0.1cm}  vacinada    &  1928  &  990   &  348 \\
\vspace{-0.1cm}  vacinal     &  1686  &  855   &  340 \\
\vspace{-0.1cm}  vacinou     &  1637  &  1012  &  335 \\
\vspace{-0.1cm}  vachina     &  1487  &  670   &  274 \\
\vspace{-0.1cm}  vacinei     &  147   &  130   &  94  \\
\vspace{-0.1cm}  v4c1n4      &  14    &  8     &  8   \\
\bottomrule
\end{tabular}
}
\caption{Key words for the vaccine filter.}
\label{tab:similaridade}
\vspace{-0.5cm}
\end{table}

After filtering, we obtained 84{,}640 vaccination-related messages from 15{,}148 users. Table~\ref{tab:similaridade} summarizes keyword coverage, including alternative terms such as ``vachina'' (1{,}487 occurrences) and ``v4c1n4'' (14 occurrences), which indicate pejorative and encoded references commonly associated with ideological and conspiratorial discourse. Although the broader corpus is largely political, those large public WhatsApp groups also served as key communication spaces for many other topics, such as health issues, during the analyzed period. While this dataset represents one of the largest known WhatsApp corpora related to vaccines, the absence of official WhatsApp-access mechanisms prevents precise claims about population representativeness.\looseness=-1

Finally, to ensure data quality and reduce redundancy, we applied semantic deduplication. Message embeddings were generated using \texttt{SentenceTransformers}~\cite{reimers-2019-sentence-bert} and indexed with \texttt{Faiss}~\cite{douze2024faiss} for nearest-neighbor similarity search. Empirically, most near-duplicates appeared below a distance threshold of 0.02, which we adopted to group equivalent messages, retaining only the first instance in each cluster. This step removed 5.18\% of messages, resulting in a final dataset of 80{,}257 unique messages from 14{,}322 users across 932 public WhatsApp groups.\looseness=-1

\subsection{Expert Medical Annotation}

Message annotation was conducted by four medical professionals using a progressive, iterative, protocol-refinement process to ensure consistent criteria and stable decision-making. Because the evaluators had no prior experience with computational labeling tasks, the procedure began with a small pilot phase to align expectations, clarify the study scope, and define what would constitute health misinformation in this context.\looseness=-1

In the first stage, each evaluator labeled 50 randomly selected messages. The resulting annotations were analyzed to assess agreement and identify sources of ambiguity, which led to targeted refinements in the guidelines. A second round of 50 messages was then completed, followed by an intermediate stage with 300 messages to evaluate whether consistency was maintained at a larger scale. At the end of this stage, Fleiss’ Kappa reached 0.65, indicating ``substantial agreement'' according to \cite{Landis1977-fx} and validating the protocol for full-scale annotation.\looseness=-1

In the final phase, the same four experts annotated 950 randomly selected messages, maintaining thematic diversity and natural temporal distribution. The result is a dataset in CSV format containing approximately 30\% misinformation and 70\% non-misinformation, with 950 WhatsApp messages consistently labeled by medical specialists, serving as the gold-standard reference for the analyses and classification experiments presented in this work. Each entry includes the message text, all four annotators' labels, anonymized group and sender metadata, temporal information, indicators of whether the message is forwarded or quoted, and the sender's country or area codes.\looseness=-1

\subsection{Limitations}

Although \textbf{WhaVax} provides a rare, medically validated view of vaccine discourse within encrypted messaging environments, some limitations remain. First, the dataset is derived from Brazilian public WhatsApp groups, which represent only a portion of the broader messaging ecosystem. As a result, it may overrepresent politically engaged, highly vocal communities and underrepresent private, familial, or less politically exposed conversations. Temporal coverage, platform-specific features, and evolving sociopolitical contexts may also limit generalization to other time periods, populations, or countries.\looseness=-1

Second, despite expert curation, annotation inevitably involves judgment and may introduce biases, especially in borderline or ambiguous messages, even though it yielded a final Fleiss’ Kappa value of 0.621, which is considered good. These cases reflect the inherent uncertainty of real-world health communication and the challenges of interpreting irony, cultural references, or implicitly framed narratives, which can still cause interpretive skew. Even with remaining gray areas, the sustained involvement of qualified medical professionals throughout dataset conception and validation substantially reduces clinical misinterpretation. Beyond strengthening scientific rigor, this participation is increasingly required by emerging regulations for medical datasets, reinforcing ethical robustness, reliability, and legal compliance.\looseness=-1

\section{Dataset Characterization} \label{sec:results}
This section characterizes the annotated dataset by describing the label distribution, expert agreement dynamics, and the presence of ambiguous instances that challenge automatic classification. Together, these analyses provide a behavioral perspective on the corpus.\looseness=-1

\subsection{Annotation and Agreement Pattern Distribution}
\begin{figure}[t]
    \centering
    \includegraphics[width=0.8\linewidth]{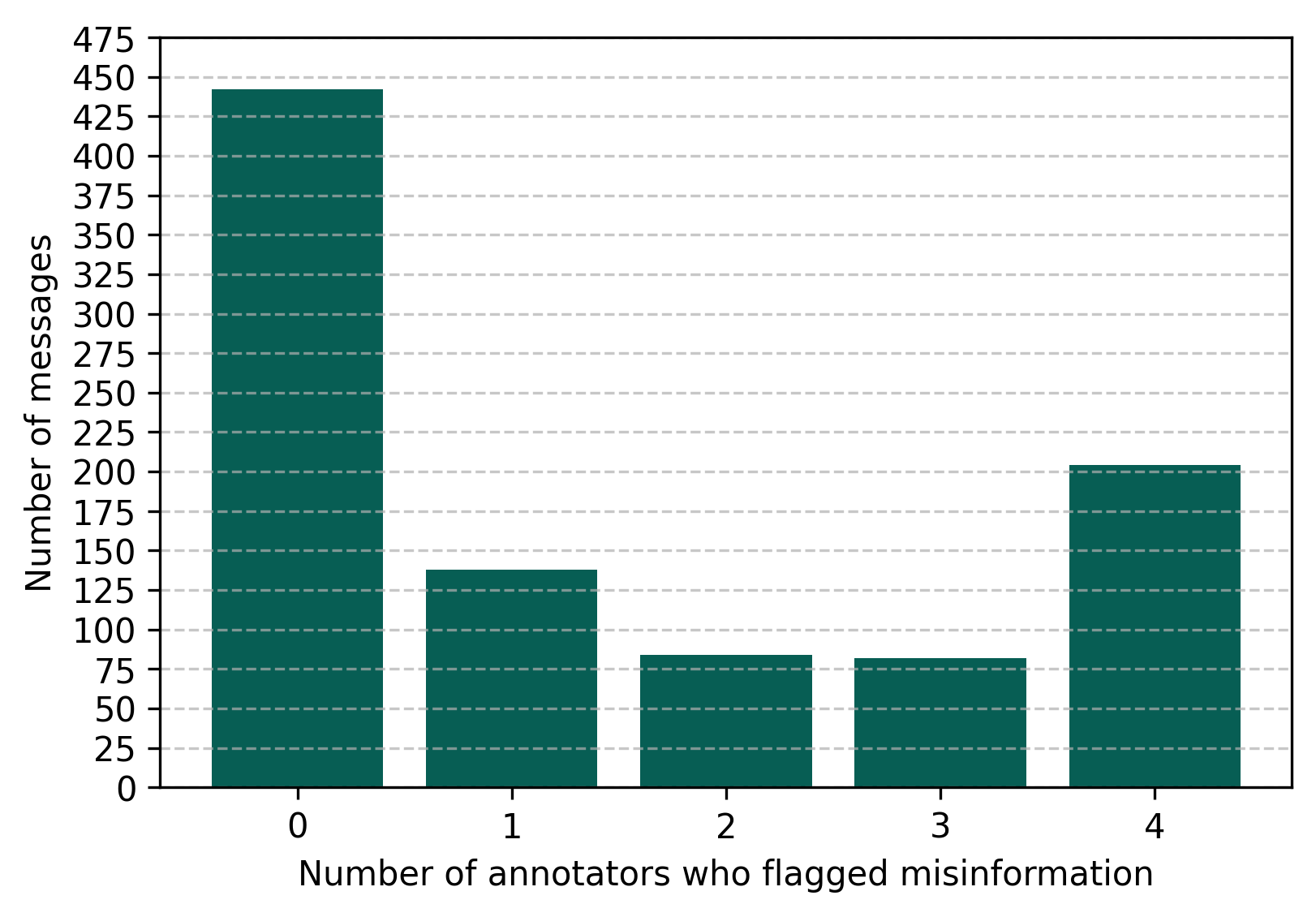}

    \vspace{-0.3cm}
    \caption{Message distribution of annotators agreements.}
    \label{fig:concordavaliadores}
    \vspace{-0.3cm}
\end{figure}

Of the 950 messages evaluated, 286 ($\approx30$\%) were labeled as misinformation under the majority criterion (at least three out of four votes), while the remaining 664 messages were classified as non-misinformation, resulting in considerable imbalance (skewness). As shown in Figure~\ref{fig:concordavaliadores}, agreement varies substantially: 442 messages were unanimously classified as non-misinformation and 204 as misinformation, indicating a sizable subset of clearly identifiable cases.

\subsection{Linguistic and Structural Messages Analysis}
\begin{figure}[t]
    \centering
    \includegraphics[width=0.8\linewidth]{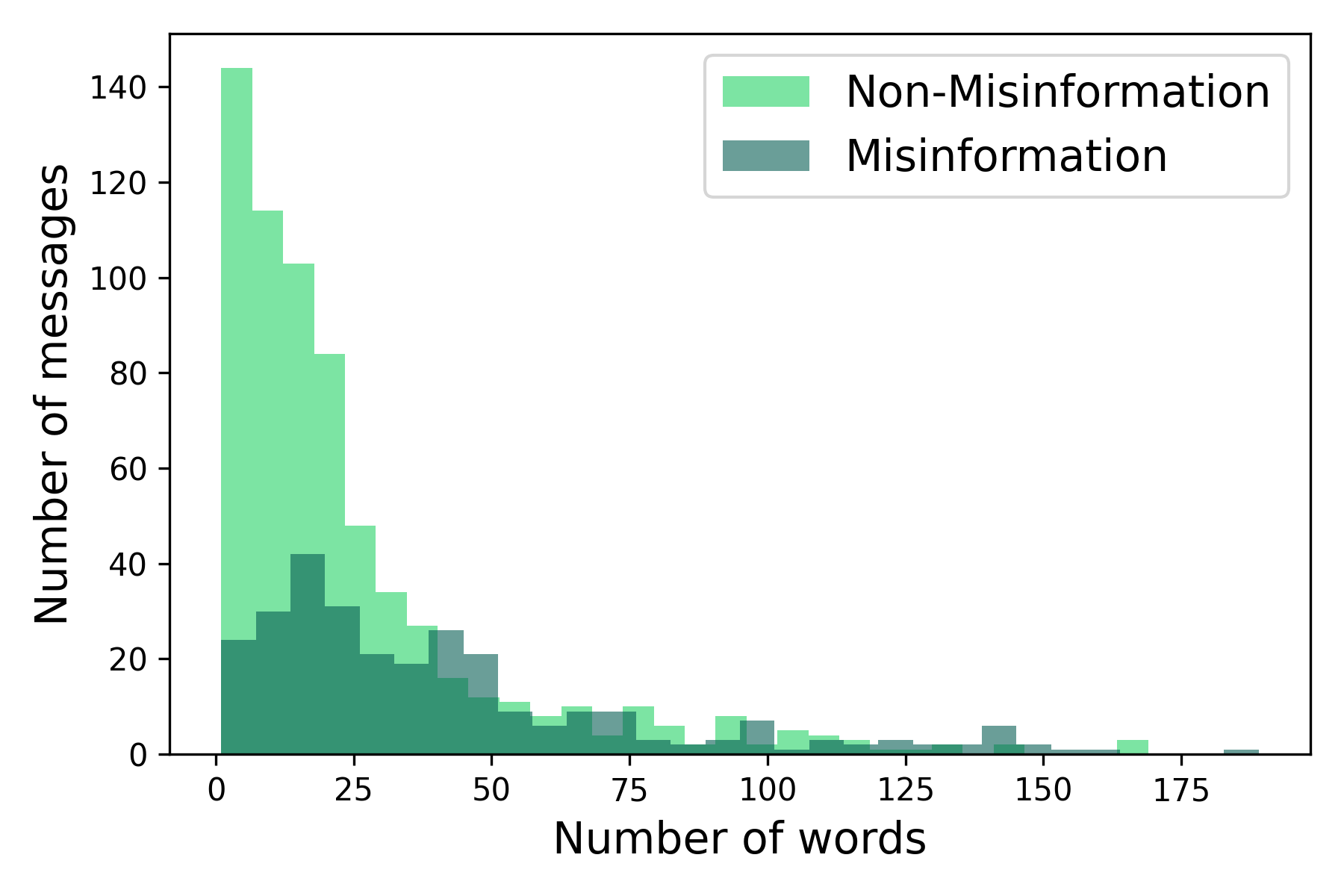}

    \vspace{-0.3cm}
    \caption{Message size distribution.}
    \label{fig:distrisizemsg}
    \vspace{-0.3cm}
\end{figure}

Further characterization of the dataset was based on the textual properties of the messages. Clear differences emerge between misinformation and non-misinformation content. As shown in Figure~\ref{fig:distrisizemsg}, misinformation messages are noticeably longer, containing on average about 290 characters and nearly 40 words, whereas non-misinformation messages contain roughly 211 characters and 22 words.\looseness=-1

These findings indicate that misinformation tends to be more elaborate, often relying on extended explanations, causal narratives, or lists of alleged adverse effects. Such verbosity may serve a persuasive role by creating an impression of credibility or overwhelming readers with excessive detail, potentially masking incorrect claims.\looseness=-1

In addition to message length, notable stylistic differences appear in the use of expressive punctuation. As shown in Table \ref{tab:dadostextuais}, misinformation messages contain substantially more exclamation points and question marks, suggesting a more emotional or rhetorical discourse style. This pattern aligns with engagement strategies commonly observed in misinformation, where emotional appeal is used to enhance persuasion or introduce doubt. A similar trend is seen with emojis: misinformation messages contain approximately four times as many emojis as non-misinformation content, reinforcing affective cues and amplifying the intended emotional impact on readers.\looseness=-1

\begin{table}[ht]
    \vspace{-0.25cm}
    \centering
    \resizebox{\linewidth}{!}{%
        \begin{tabular}{@{}lrccccc@{}}
            \toprule
            \textbf{Class} & \textbf{\#Characters} & \textbf{\#Words} & \textbf{\#Exclamations} & \textbf{\#Question Marks} & \textbf{\#Capital letters} \\ \midrule
            Non Misinformation                & 210.92 & 22.43 & 0.22 & 0.29 & 12.57\\
            Misinformation                & 290.39 & 39.62 & 0.59 & 0.46 & 35.67\\ \bottomrule
        \end{tabular}%
    }
    \vspace{-0.3cm}
    \caption{Textual characteristics.}
    \label{tab:dadostextuais}
    \vspace{-0.3cm}
\end{table}

The use of capital letters also emerges as a relevant discriminative feature. Messages labeled as misinformation contain, on average, nearly three times more capitalized text than informational messages. This pattern suggests deliberate emphasis strategies, likely intended to attract attention or convey a sense of urgency, further reinforcing the emotional tone commonly associated with misinformation content.\looseness=-1

\subsection{Lexical Patterns and N-gram Analysis}

Analysis of the most frequent n-grams also highlights semantic distinctions between classes. In misinformation messages, terms such as ``mRNA’’, ``efeitos’’, ``morte’’, ``não’’, and ``tomar’’ are more frequent, usually associated with narratives of risk, fear, or doubt towards treatments and vaccines. This recurrent usage emphasizes potential negative outcomes and distrust of medical interventions. In contrast, non-misinformation messages feature n-grams linked to institutional sources, public policies, and health campaigns—such as ``vacinação’’, ``governo’’, and ``contra covid’’—suggesting a more informative discourse aligned with official communications and public health guidance.\looseness=-1

These lexical distinctions reinforce the idea that health misinformation is conveyed not only through factually incorrect claims but also through distinctive linguistic patterns, which classification models can systematically identify.\looseness=-1

\subsection{Group-Level Distribution of Misinformation}

\begin{figure}[t]
    \centering
    \includegraphics[width=0.8\linewidth]{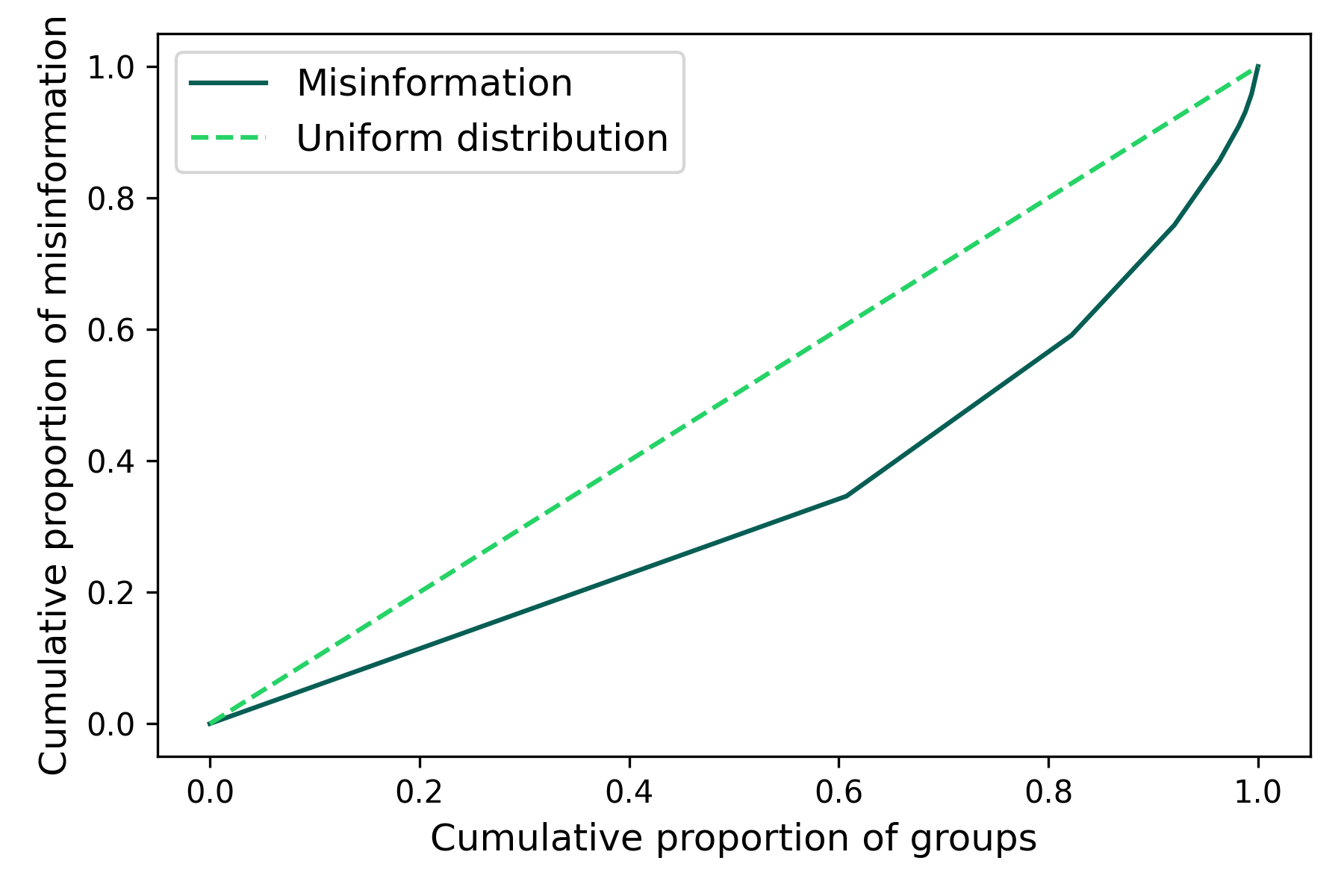}

    \vspace{-0.3cm}
    \caption{Concentration of misinformation by group.}
    \label{fig:concenmis}
    \vspace{-0.3cm}
\end{figure}

Group-level analysis shows that misinformation is unevenly distributed across communities, as illustrated in Figure~\ref{fig:concenmis}. Some groups have much higher concentrations of misinformation, with proportions exceeding 60\% and, in a few cases, approaching 100\%. This indicates highly polarized groups where misinformation prevails.\looseness=-1

Conversely, a significant portion of messages displayed ambiguity; specifically, 84 messages resulted in tied evaluations among the four experts, representing approximately 8.8\% of the dataset. These cases highlight the challenges of assessing health misinformation, particularly in short or opinion-driven messages. Following a conservative strategy, these cases were labeled as non-misinformation. Additionally, no clear linear relationship exists between forwarding and misinformation, indicating that sharing depends more on group-specific social and contextual factors than on content alone.

\subsection{Temporal Analysis}

\begin{figure}[t]
    \centering
    \includegraphics[width=0.9\linewidth]{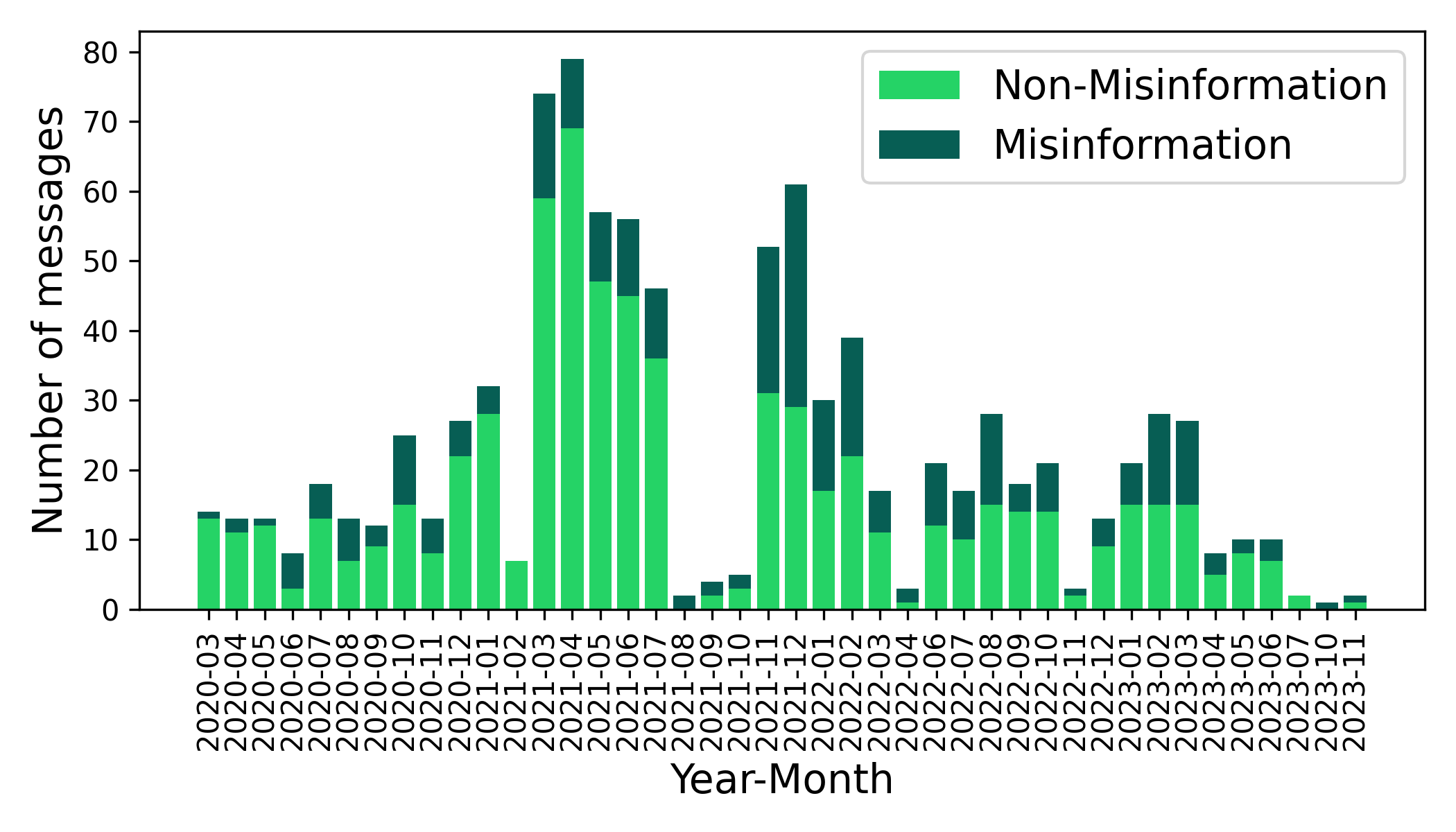}

    \vspace{-0.3cm}
    \caption{Temporal distribution of messages.}
    \label{fig:tempdistri}
    \vspace{-0.3cm}
\end{figure}

Figure~\ref{fig:tempdistri} presents the monthly temporal distribution of messages from 2020 to 2024, distinguishing between misinformation and non-misinformation content. Instead of focusing on specific peaks, this analysis evaluates dataset coverage and the coexistence of informative and misleading content over time. Misinformation is consistently present throughout the entire period, including months with lower activity, indicating its persistent nature. The dataset offers broad temporal coverage, allowing computational analyses to include messages from different phases and discursive contexts of the pandemic, which helps mitigate temporal bias and enhances the robustness of subsequent experiments.\looseness=-1

Two major activity peaks emerged in 2021. The first and most pronounced occurred in March 2021, coinciding with the arrival of COVID-19 vaccines in Brazil and the widely publicized first administration~\cite{primeiravacina,primeiravacinaSP}. This marked a discursive turning point on WhatsApp, with an intensified presence of supportive, hesitant, and rejection narratives regarding vaccination. A second peak appeared at the end of 2021, aligned with the surge in COVID-19 mortality driven by the Omicron variant~\cite{retrospectiva} and ANVISA’s authorization of Pfizer’s vaccine for children aged 5–11~\cite{vacinapfizer}, which reignited debate marked by increased misinformation, moral panic, and conspiratorial narratives~\cite{De_Albuquerque2022-ro}.\looseness=-1

\section{Classification Experiments – Approaches, Experimental Setup and Results}

We frame our problem as a binary text classification task, where each of the 950 expert-annotated WhatsApp messages is labeled as either \textit{misinformation} or \textit{non-misinformation}. The distribution is naturally imbalanced (approximately 70\% non-misinformation and 30\% misinformation), reflecting real-world prevalence and presenting additional challenges due to limited data, informal writing, and highly contextual conversational language.\looseness=-1

To thoroughly investigate model behavior under these realistic constraints, we evaluate three families of approaches: (i) classical machine learning models, (ii) fine-tuned Transformer-based Small Language Models (SLMs), and (iii) Large Language Models (LLMs) using in-context learning (ICL). Classical supervised classifiers—Support Vector Machines (SVM), Logistic Regression, Random Forest, Multilayer Perceptron (MLP), and XGBoost—were tested with different embedding strategies to assess the impact of representation quality on performance \cite{Kowsari_2019}.\looseness=-1

For SLMs, we fine-tuned Portuguese and biomedical-oriented Transformers, including BERTimbau~\cite{souza2020bertimbau}, RoBERTa~\cite{roberta}, BERTuguês~\cite{Zago2024bertugues}, and BioBERTpt~\cite{schneider-etal-2020-biobertpt}. We optimized hyperparameters (learning rate, weight decay) using nested cross-validation to maximize performance and mitigate overfitting.\looseness=-1 

We then examine LLMs in zero- and few-shot settings, focusing on in-context classification without parameter training. We evaluate open-source models such as LLaMA 3.1 (8B), LLaMA 3.2 (3B), DeepSeek v3.2 (685B)~\cite{deepseekai2025deepseekv32}, and Qwen 3 (30B)~\cite{qwen3technicalreport}, enabling analysis of model size, architecture, and language alignment effects. For LLaMA 3.1, we also vary the number of in-context examples to assess sensitivity to prompt conditioning. Finally, we include proprietary models accessed via API (GPT-5.1 and GPT-5.2\footnote{openai.com/api/}), evaluated in few-shot mode to compare open and proprietary systems and to explore whether more recent knowledge grounding affects performance.\looseness=-1

This multifaceted evaluation enables a direct, fair comparison across paradigms, highlighting when simple models suffice, when domain-adapted Transformers help, and when modern LLMs provide advantages.\looseness=-1
\vspace{-0.1cm}

\subsection{Experimental Setup}
We use 5-fold stratified cross-validation to maintain class balance across folds and improve the robustness of estimates \cite{raschka2020modelevaluationmodelselection}. To further reduce sampling variance, we repeat the entire cross-validation process eight times with controlled shuffling, generating distributions of results rather than single-point estimates. For each model, we report the mean and standard deviation across runs.\looseness=-1

We compute accuracy, precision and recall for each class, along with macro-F1 as the primary comparison metric, since it assigns equal weight to misinformation and non-misinformation—an essential property in imbalanced scenarios. Statistical significance between competing models is assessed using the paired Wilcoxon test over cross-validation runs, which avoid normality assumptions while ensuring reliable comparative claims. This evaluation framework provides a rigorous and comparable basis for assessing classical models, fine-tuned SLMs, and in-context LLMs under the same conditions. Table~\ref{tab:metricamodelo} summarizes performance using macro-F1 as the main criterion, with model families discussed further in the following sections.\looseness=-1

\begin{table}[t]
    \vspace{-0.25cm}
    
    \resizebox{0.9\linewidth}{!}{%
        \begin{tabular}{@{}lrcc@{}}
            \toprule
            \textbf{Model} & \textbf{Average macro-F1} \\ \midrule
            \textbf{Bertimbau Embedding}\\
            Support Vector Machine      & 0.724 ± 0.030\\
            Logistic Regression                 & 0.747 ± 0.036\\
            Random Forest                    & 0.656 ± 0.041\\
            Multi-Layer Perceptron          & 0.738 ± 0.031\\
            XGBOOST       & 0.687 ± 0.034\\
            \textbf{Qwen8b Embedding}\\
            Support Vector Machine      & 0.784 ± 0.025\\
            Logistic Regression        & \textbf{0.791 ± 0.026}\\
            Random Forest                    & 0.663 ± 0.035\\
            Multi-Layer Perceptron          & \textbf{0.789 ± 0.026}\\
            XGBOOST       & 0.729 ± 0.035\\
            \hline\hline
            BerTimbau & 0.712 ± 0.101\\
            Roberta        & 0.580 ± 0.116\\
            Bertugues             & 0.660 ± 0.056\\
            BioBertpt            & 0.634 ± 0.102\\
            \hline\hline
            Llama 3.1 8B (ICL Mode - 8 examples)     & 0.734 ± 0.028\\
            Llama 3.2 3B (ICL Mode - 8 examples)    & 0.503 ± 0.031\\
            Qwen 30B (ICL Mode - 8 examples)        & 0.383 ± 0.027\\
            DeepSeek-V3.2 685B (ICL Mode - 8 examples) & 0.744 ± 0.024\\
            GPT 5.1 (ICL Mode - 8 examples)  & \textbf{0.786 ± 0.026}\\
            GPT 5.2 (ICL Mode - 8 examples)   & 0.780 ± 0.023\\ \bottomrule
        \end{tabular}%
    }
    \vspace{-0.3cm}
    \caption{Macro F1 - Classifications models.}
    \label{tab:metricamodelo}
    \vspace{-0.3cm}
\end{table}

\subsection{Classical Machine Learning Models with Contextual Embeddings}

Messages were represented using embeddings from pre-trained language models to provide classical classifiers with dense semantic encodings. BERTimbau embeddings were initially used, computed as the mean of the last hidden-layer token representations, yielding strong baseline results with Logistic Regression. In a second phase, all models were re-evaluated with Qwen (8B) embeddings under the same pipeline, enabling a controlled assessment of representation effects. Classical models show stable performance, with Logistic Regression and Qwen (8B) achieving the best result (macro-F1 = 0.791). The comparison confirms that richer embeddings consistently outperform BERTimbau (macro-F1 = 0.747), highlighting the central role of representation quality in handling the informal and noisy language of WhatsApp messages.

The gains achieved by Qwen (8B) extend beyond Logistic Regression, delivering consistent macro-F1 improvements across all classical models compared to BERTimbau, including SVM (0.724 → 0.784) and MLP (0.738 → 0.789). Even models that typically struggle with dense, high-dimensional vectors, such as Random Forest and XGBoost, show modest improvements, reinforcing that embedding quality often has a greater influence than classifier architecture in lexically complex settings. Qwen’s superior performance likely results from its larger scale and broader pre-training, which better capture semantic nuance, informal language, and implicit references common in WhatsApp messages. These results highlight that robust embeddings can compensate for simpler classifiers, underscoring the central role of textual representation in misinformation detection.\looseness=-1

\vspace{-0.1cm}
\subsection{Fine-Tuned Small Language Models (SLMs)}
\vspace{-0.1cm}

Fine-tuned Small Language Models (SLMs) show more heterogeneous performance. BERTimbau achieves a macro-F1 of 0.712, comparable to classical models but with noticeably higher variance, indicating sensitivity to training splits and hyperparameters under data scarcity. Other SLMs, including BERTuguês, BioBERTpt, and especially RoBERTa, perform below expectations. This likely results from a mismatch between their pre-training domains and the characteristics of our dataset, which features informal writing, abbreviations, typos, and conversational phrasing typical of instant messaging. The limited sample may also be insufficient to adapt these models for further generalization. Fine-tuned SLMs show variable performance and depend heavily on domain alignment and data availability. In our setting, simpler models with strong embeddings were more reliable.\looseness=-1

\begin{figure}[t]
    \centering
    \includegraphics[width=0.8\linewidth]{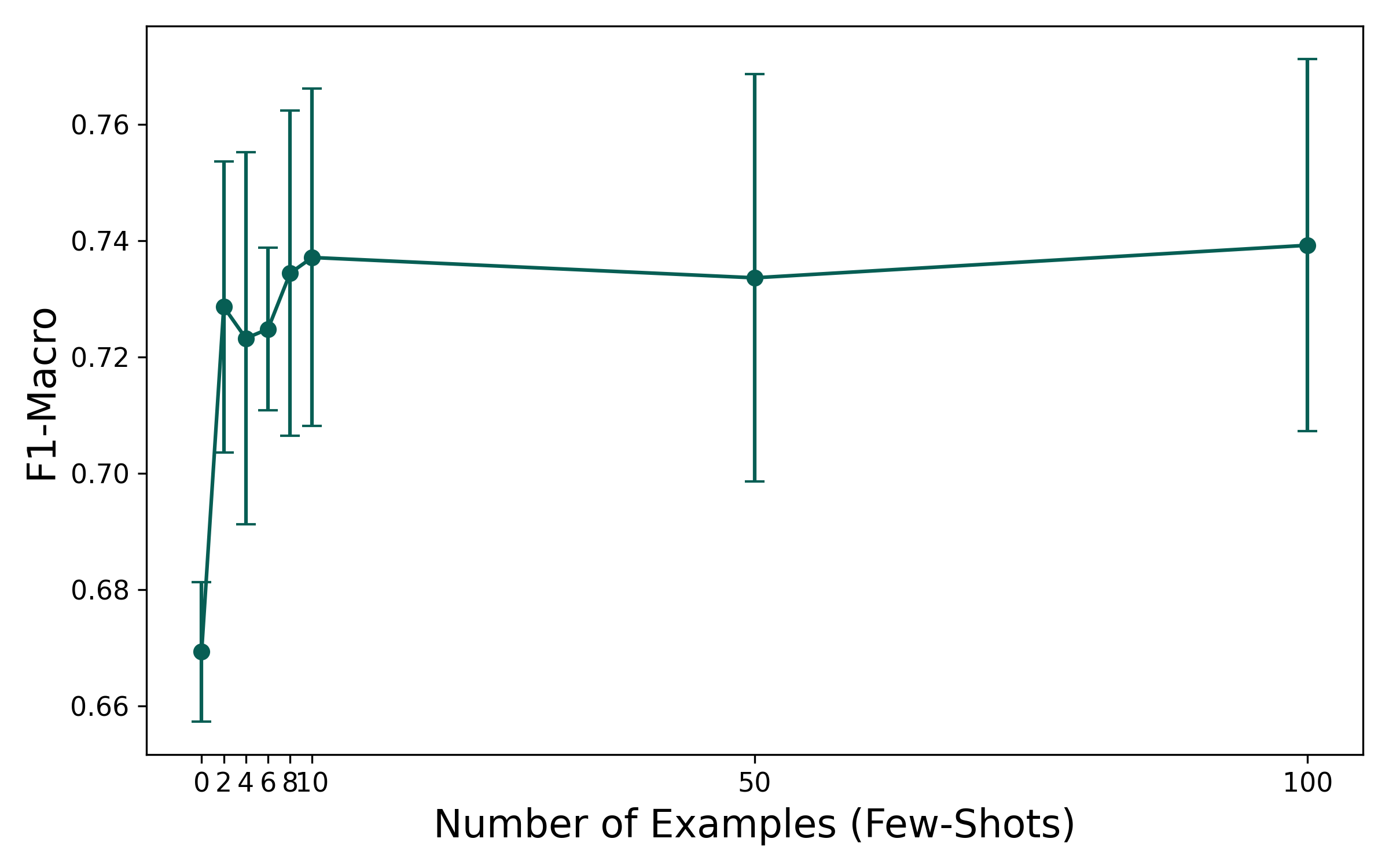}

    \vspace{-0.3cm}
    \caption{Impact of few-shots on performance.}
    \label{fig:fewperformance}
    \vspace{-0.3cm}
\end{figure}

\subsection{Large Language Models with In-Context Learning}

\subsubsection{Few-Shot Behavior and Context Sensitivity.}

We first analyze the few-shot\footnote{Few-shot learning is a type of in-context learning (ICL) in which typically a few  examples (2-50) are used in the prompt.} behavior of LLaMA 3.1 (8B), which serves as our reference model. Figure~\ref{fig:fewperformance} shows its macro-F1 performance as a function of the number of examples provided in the prompt. The x-axis is shown on a logarithmic scale to better capture performance variations in low-context regimes, where changes tend to be most informative. A clear gain is observed when moving from zero-shot to few-shot configurations: even a small number of examples substantially improves performance, indicating that the model quickly internalizes discriminative cues for the task. However, this improvement plateaus around eight examples, after which additional context yields only minor oscillations within the variance margin.\looseness=-1

This pattern suggests that most of the signal for distinguishing misinformation is captured in early examples, with diminishing returns as context increases. Based on this, we use eight shots for all LLM experiments to balance performance, cost, and comparability. The prompt used for Few-Shot consisted of an instruction to the models and eight already labeled examples:\looseness=-1

\begin{table*}[htbp]
\centering
\scriptsize
\begin{tabularx}{\textwidth}{X|X|c|c}
\hline
\textbf{Portuguese} & \textbf{English (Translation)} & \textbf{Correct Label} & \textbf{N Incorrect Models
}\\
\hline
     CENTENAS DE MILHARES ESTÃO MORTOS E  ENTERRADOS PORQUE O \censor{BOZO} RECUSOU-SE A COMPRAR VACINAS! POR CAUSA DISSO A CONTAMINAÇÃO AUMENTOU E EXPLODIU!!! \censor{BOZO} É UM ASSASSINO MILICIANO GENOCIDA! &  HUNDREDS OF THOUSANDS ARE DEAD AND BURIED BECAUSE \censor{BOLSONARO} REFUSED TO BUY VACCINES! BECAUSE OF THIS, THE CONTAMINATION INCREASED AND EXPLODED!!! \censor{BOLSONARO} IS A GENOCIDAL MILITIA MURDERER! & 0 & 18 \\
\hline
     BÉLGICA EM FÚRIA  Bruxelas sob domínio da Guerra  Os holandeses e agora os Belgas. A humanidade está saindo do ostracismo das manifestações pacíficas e idiotas para a verdadeira reação contra a tirania sanitária.   O alvo são as cabeças dos políticos, dos governantes e também dos jornalistas corruptos que promovem a tirania das vacinas genocidas e patrocinam a ditadura dos passaportes sanitários.  Parece que a fúria foi despertada. Agora a Humanidade está acordando para a realidade: de que não há conquistas sem guerras.  Bruxelas amanheceu debaixo da ira da população que atacou o sistema com pedras, porretes e muita gasolina. A fúria e a raiva chegarão nos responsáveis, que serão caçados, trucidados e enviados de volta para o inferno.  NÃO RECUEM! NÃO SE DOBREM! ARREBEBENTEM COM TUDO! &  BELGIUM IN FURY Brussels under the dominion of War The Dutch and now the Belgians. Humanity is emerging from the ostracism of peaceful and idiotic demonstrations into a true reaction against sanitary tyranny. The target is the heads of politicians, rulers, and also corrupt journalists who promote the tyranny of genocidal vaccines and sponsor the dictatorship of health passports. It seems that fury has been awakened. Now Humanity is waking up to the reality: that there are no conquests without wars. Brussels awoke under the wrath of the population that attacked the system with stones, clubs, and plenty of gasoline. The fury and rage will reach those responsible, who will be hunted down, slaughtered, and sent back to hell. DO NOT RETREAT! DO NOT BEND! BREAK EVERYTHING DOWN! & 0 & 17
\\
\hline
     Pessoal, percam uma horinha de vocês e vejam esta reportagem.   É estarrecedora e muito esclarecedora.  São dois cientistas, sendo que um deles é umas maiores sumidades sobre vacinas, responsável pelo design de vacinas em grandes órgãos públicos e empresas privadas.  Veja o que ele fala sobre a vacinação.  É muito esclarecedor. &  Folks, take an hour of your time and watch this report. It's shocking and very enlightening. It features two scientists, one of whom is a leading authority on vaccines, responsible for vaccine design in major public agencies and private companies. Watch what he says about vaccination. It's very insightful. & 0 & 17
\\
\hline
     China admite que eficácia de suas vacinas contra Covid-19 não é alta &  China admits that the effectiveness of its Covid-19 vaccines is not high. & 1 & 16
 \\
\hline
     Vai saber se foi alguma vacina que deu errado &  Who knows if it was some vaccine that went wrong? & 1 & 16
\\    

\hline
\end{tabularx}
\caption{Examples incorrectly classified by most models.}
\label{tab:examplesmessageserror}
\end{table*}

\begin{quote}
\textbf{English translation of in-context learning prompt}
\tiny
\begin{verbatim}
You are a health misinformation 
classifier.

Classify messages as:
0 = Non-misinformation or
1 = Is misinformation

Here are some examples:
{few_shot_examples}

Now classify the following message,
respond strictly with 0 or 1,
do not respond in any other way.

Text: [Message]

Answer:
\end{verbatim}
\end{quote}

\subsubsection{Few-Shot Behavior and Context Sensitivity.}

LLMs evaluated under this standardized few-shot regime present heterogeneous outcomes. LLaMA 3.1 (8B) and DeepSeek-V3.2 (685B) achieve competitive macro-F1 scores near 0.74, comparable to the best classical models. These results indicate that, even without supervised fine-tuning, LLMs can effectively learn the task from limited in-context supervision.\looseness=-1

In contrast, smaller or less linguistically aligned models, such as LLaMA 3.2 (3B) and Qwen (30B), perform substantially worse. This indicates that success in in-context learning depends not only on scale, but also on pre-training quality, language coverage, and instruction-following capability. Overall, our findings confirm that in-context learning is powerful but not universally reliable, with performance sensitive to model design and task domain alignment.\looseness=-1

\subsubsection{API-Based in-context Learning Models.}

\begin{table}[htbp]
    \vspace{-0.25cm}
    \centering
    \resizebox{\linewidth}{!}{%
        \begin{tabular}{@{}l|c|c|c|c|c@{}}
            \toprule
            \textbf{Model} & \textbf{Accuracy} & \textbf{Precision (0)} & \textbf{Recall (0)} & \textbf{Precision (1)} & \textbf{Recall(1)} \\ \midrule
            LR        & 0.825 ± 0.02 & 0.875 ± 0.02 & 0.876 ± 0.02 & 0.711 ± 0.04 & 0.707 ± 0.06  \\
            MLP           & 0.827 ± 0.02 & 0.865 ± 0.02 & 0.893 ± 0.02 & 0.731 ± 0.04 & 0.674 ± 0.05\\
            GPT 5.1   & 0.798 ± 0.03 & 0.964 ± 0.02 & 0.739 ± 0.04 & 0.607 ± 0.03 & 0.935 ± 0.03\\ \bottomrule
        \end{tabular}%
    }
    \vspace{-0.2cm}
    \caption{Metrics for the Top 3 models.}
    \label{tab:metricabestmodels}
    \vspace{-0.3cm}
\end{table}

Finally, GPT-based models 
rank among the strongest performers. GPT-5.1 reaches a macro-F1 of 0.786, and GPT-5.2 achieves 0.780, both with low variance, statistically matching the best classical approach (Logistic Regression with Qwen embeddings). These results position GPT models as highly effective, stable alternatives for misinformation detection in Portuguese WhatsApp messages without fine-tuning.\looseness=-1

Their advantage likely stems from large-scale pre-training, strong instruction-following, and broad prior exposure to health-related data. Interestingly, the small gap between GPT-5.1 and GPT-5.2 suggests that temporal knowledge updates play a smaller role, while general reasoning and semantic understanding appear to be more important.
\looseness=-1

\subsection{In-depth Analysis of the  Best Models}

Table~\ref{tab:metricabestmodels} presents the effectiveness of the three best-performing models – LR, MLP (both using Qwen embeddings), and GPT-5.1—broken down into Overall Accuracy (MicroF1), and Precision and Recall for non-misinformation (Class 0) and misinformation (Class 1). The table shows that all three models achieve very similar Macro-F1 scores, with GPT-5.1 having a slight disadvantage. However, their behavior differs in important ways. LR and MLP, both using contextual embeddings, exhibit similar patterns: they achieve higher precision for non-misinformation but are less effective at identifying misinformation, with precision and recall for this class in the 0.67–0.73 range. In contrast, GPT-5.1 demonstrates a different profile. It excels in precision for non-misinformation and, importantly, in recall for misinformation, meaning it identifies most misinformation cases while accepting a higher rate of false positives. In summary, although overall effectiveness is comparable, LR and MLP provide more conservative and balanced decisions, whereas GPT-5.1 prioritizes high recall for misinformation, making it suitable for screening scenarios where missing misinformation is unacceptable.\looseness=-1

\subsection{Model Error Analysis}

Table~\ref{tab:examplesmessageserror} shows common misclassification patterns across models. The first two cases, misclassified by 18 and 17 out of 20 models, contain highly emotional and politicized language. Although medical experts labeled them as non-misinformation, most models flagged them as misinformation, likely because they associate inflammatory rhetoric and pandemic narratives with false content. This suggests confusion between hate speech or extremism and factual misinformation.\looseness=-1

A second example appears in messages that appeal to scientific authority or invite reflection. Even without explicit false claims, references to ``scientists'' and ``revelations'' triggered misinformation predictions. Models struggle to distinguish skepticism from actual incorrect information. The bottom examples show the opposite difficulty: short, ambiguous statements – sometimes used manipulatively – were often misclassified as non-misinformation. These involve insinuation, framing, or selective facts that subtly foster distrust, which models frequently fail to detect.\looseness=-1

Overall, these errors reinforce that misinformation detection goes beyond standard text classification and requires sensitivity to discourse, pragmatics, and context, dimensions that current models still struggle to capture.\looseness=-1

\section{Conclusion and Future Work} \label{sec:conclusion}

This paper introduces \textbf{WhaVax} -- a new expert-annotated dataset of vaccine-related WhatsApp messages, providing one of the first medically validated resources that capture how health misinformation circulates within encrypted messaging environments. Through rigorous filtering and a structured multi-stage labeling process with medical professionals, our dataset was derived from large-scale WhatsApp public group data collection, demonstrating the scientific value and feasibility of such approaches.

Our analyses revealed distinct linguistic, structural, lexical, temporal, and group-level traits of misinformation, and highlighted a meaningful subset of inherently ambiguous messages that reflect the complexity of real-world health communication. Classical models with strong embeddings remain competitive, fine-tuned SLMs depend on domain alignment and data availability, and zero-/few-shot LLMs perform well without supervision, although with substantial variation across architectures. Collectively, these results provide an empirically grounded resource and key insights for misinformation detection in private messaging platforms. We hope that releasing our dataset and methodology will foster new research directions and help address the scarcity of annotated data in this domain.

\subsubsection{Modeling Annotation Disagreement and Ambiguity.}

Given that the dataset includes multiple annotations per message, there is a valuable opportunity to study disagreement among health misinformation expert annotators. This enables research on ambiguity, borderline cases, and subjectivity in issues that are especially important in health-related contexts, where opinion, uncertainty, and emerging evidence often intersect. Such analyses can inform the design of more robust annotation schemes and evaluation protocols.

\subsubsection{Informing Public Health Interventions and Policy.}

Insights from this dataset may inform public health campaigns by highlighting recurring narratives, misconceptions, and sources of confusion about vaccines. Although not intended for direct operational use, the dataset can support evidence-based discussions on how health authorities and fact-checkers might better address misinformation circulating in instant messaging ecosystems.

 \section{Ethics Statement and Dataset Availability}

This study was conducted in compliance with established biomedical research ethics standards and was reviewed and approved by an independent Research Ethics Committee (IRB/CEP equivalent). The protocol was approved under the Brazilian national ethics system (Plataforma Brasil).\looseness=-1 

All data collection, processing, and release protocols were designed to minimize risk and protect participant privacy. Only de-identified 
data are included in the dataset, and access is restricted to research and educational purposes.\looseness=-1

Importantly, the study included active participation of qualified medical professionals throughout dataset conception, curation, and validation. Their involvement ensured clinical relevance, ethical rigor in handling sensitive health-related data, and alignment with evolving regulatory expectations that increasingly require medical oversight in the development and dissemination of healthcare datasets.\looseness=-1

Last, the dataset build in this paper is publicly available here: \url{https://zenodo.org/records/18190030}.

\section{Acknowledgments}
This work was funded by the Center for Innovation and Artificial Intelligence in Health (CI-IA Saúde), with financial support from the São Paulo Research Foundation (FAPESP), grant no. 2020/09866-4; the Minas Gerais Research Foundation (FAPEMIG), grant no. PPE-00030-21; and UNIMED Belo Horizonte. It was also supported by the National Institute of Science and Technology in Responsible Artificial Intelligence for Computational Linguistics, Information Treatment, and Dissemination (INCT-TILDIAR), funded by the Brazilian National Council for Scientific and Technological Development (CNPq), grant no. 408490/2024-1. Additional support was provided by FAPEMIG, including grants APQ-04803-25 and APQ-03115-24.

{\small
\bibliography{aaai2026}

@article{Weinzierl_Harabagiu_2022, title={From Hesitancy Framings to Vaccine Hesitancy Profiles: A Journey of Stance, Ontological Commitments and Moral Foundations}, volume={16}, url={https://ojs.aaai.org/index.php/ICWSM/article/view/19360}, DOI={10.1609/icwsm.v16i1.19360}, abstractNote={While billions of COVID-19 vaccines have been administered, too many people remain hesitant. Twitter, with its substantial reach and daily exposure, is an excellent resource for examining how people frame their vaccine hesitancy and to uncover vaccine hesitancy profiles. In this paper we expose our processing journey from identifying Vaccine Hesitancy Framings in a collection of 9,133,471 original tweets discussing the COVID-19 vaccines, establishing their ontological commitments, annotating the Moral Foundations they imply to the automatic recognition of the stance of the tweet authors toward any of the CoVaxFrames that we have identified. When we found that 805,336 Twitter users had a stance towards some CoVaxFrames in either the 9,133,471 original tweets or their 17,346,664 retweets, we were able to derive nine different Vaccine Hesitancy Profiles of these users and to interpret these profiles based on the ontological commitments of the frames they evoked in their tweets and on value of their stance towards the evoked frames.}, number={1}, journal={Proc. of the Int. AAAI Conference on Web and Social Media}, author={Weinzierl, Maxwell A. and Harabagiu, Sanda M.}, year={2022}, month={May}, pages={1087-1097} }

@article{weinzierl-covid-glp,
	title        = {Automatic detection of COVID-19 vaccine misinformation with graph link prediction},
	author       = {Maxwell A. Weinzierl and Sanda M. Harabagiu},
	year         = 2021,
	journal      = {Journal of Biomedical Informatics},
	volume       = 124,
	pages        = 103955,
	doi          = {https://doi.org/10.1016/j.jbi.2021.103955},
	issn         = {1532-0464},
	url          = {https://www.sciencedirect.com/science/article/pii/S1532046421002847}
}

@misc{cui2020coaid,
    title={CoAID: COVID-19 Healthcare Misinformation Dataset},
    author={Limeng Cui and Dongwon Lee},
    year={2020},
    eprint={2006.00885},
    archivePrefix={arXiv},
    primaryClass={cs.SI}
}

@ARTICLE{Landis1977-fx,
  title   = "The measurement of observer agreement for categorical data",
  author  = "Landis, J R and Koch, G G",
  journal = "Biometrics",
  volume  =  33,
  number  =  1,
  pages   = "159--174",
  month   =  mar,
  year    =  1977
}

@article{Zago2024bertugues,
    title   = {BERTugues: A Novel BERT Transformer Model Pre-trained for Brazilian Portuguese},
    volume  = {45},
    url     = {https://ojs.uel.br/revistas/uel/index.php/semexatas/article/view/50630},
    DOI     = {10.5433/1679-0375.2024.v45.50630},
    journal = {Semina: Ciências Exatas e Tecnológicas},
    author  = {Mazza Zago, Ricardo and Agnoletti dos Santos Pedotti, Luciane},
    year    = {2024},
    month   = {Dec.},
    pages   = {e50630}
}

@article{deepseekai2025deepseekv32,
  title={Deepseek-v3. 2: Pushing the frontier of open large language models},
  author={Liu, Aixin and Mei, Aoxue and Lin, Bangcai and Xue, Bing and Wang, Bingxuan and Xu, Bingzheng and Wu, Bochao and Zhang, Bowei and Lin, Chaofan and Dong, Chen and others},
  journal={arXiv preprint arXiv:2512.02556},
  year={2025}
}

@misc{qwen3technicalreport,
      title={Qwen3 Technical Report}, 
      author={Qwen Team},
      year={2025},
      eprint={2505.09388},
      archivePrefix={arXiv},
      primaryClass={cs.CL},
      url={https://arxiv.org/abs/2505.09388}, 
}

@misc{raschka2020modelevaluationmodelselection,
      title={Model Evaluation, Model Selection, and Algorithm Selection in Machine Learning}, 
      author={Sebastian Raschka},
      year={2020},
      eprint={1811.12808},
      archivePrefix={arXiv},
      primaryClass={cs.LG},
      url={https://arxiv.org/abs/1811.12808}, 
}

@inproceedings{garimella2025globalpatternsviralcontent,
    title={Global Patterns of Viral Content on WhatsApp}, 
    author={Garimella, Kiran and Cintaqia, Princessa and Rojas-Constain, Juan José and Nayak, Bharat Kumar and Vashistha, Aditya}, 
  booktitle={Proc. of the Int. AAAI Conference on Web and Social Media},
  volume={19},
  pages={586--601},
    year={2025}
}

@ARTICLE{Sharma2023-dr,
  title     = "{COVID-19-associated} misinformation across the South Asian
               diaspora: Qualitative study of {WhatsApp} messages",
  author    = "Sharma, Anjana E and Khosla, Kiran and Potharaju, Kameswari and
               Mukherjea, Arnab and Sarkar, Urmimala",
  journal   = "JMIR Infodemiology",
  volume    =  3,
  number    =  1,
  pages     = "e38607",
  month     =  jan,
  year      =  2023,
  language  = "en"
}

@article{10.1093/jamia/ocab219,
    author = {Vijaykumar, Santosh and Rogerson, Daniel T and Jin, Yan and de Oliveira Costa, Mariella Silva},
    title = {Dynamics of social corrections to peers sharing COVID-19 misinformation on WhatsApp in Brazil},
    journal = {Journal of the American Medical Informatics Association},
    volume = {29},
    number = {1},
    pages = {33-42},
    year = {2021},
    month = {11},
    issn = {1527-974X},
    doi = {10.1093/jamia/ocab219},
    url = {https://doi.org/10.1093/jamia/ocab219},
    eprint = {https://academic.oup.com/jamia/article-pdf/29/1/33/41955567/ocab219.pdf},
}

@ARTICLE{De_Albuquerque2022-ro,
  title     = "Vaccination for {COVID-19} in children: Denialism or
               misinformation?",
  author    = "de Albuquerque, Tha{\'\i}s Rodrigues and Macedo, Luis Fernando
               Reis and de Oliveira, Erika Galv{\~a}o and Neto, Modesto Leite
               Rolim and de Menezes, Irwin Rose Alencar",
  journal   = "J. Pediatr. Nurs.",
  publisher = "Elsevier BV",
  volume    =  64,
  pages     = "141--142",
  month     =  may,
  year      =  2022,
  language  = "en"
}

@misc{vacinapfizer,
  author       = {{Ricardo Brito}},
  title        = {{Brazil health regulator approves Pfizer COVID-19 shot for ages 5 to 11}},
  howpublished = {Reuters, https://www.reuters.com/world/americas/brazil-health-regulator-approves-pfizer-covid-19-shot-ages-5-11-2021-12-16/},
  year         = {2021},
 note = {Accessed on 01/09/2026}
}

@misc{retrospectiva,
  author       = {Eduardo Simões},
  title        = {{Brazil on alert after third case of Omicron variant}},
  howpublished = {Reuters, https://www.reuters.com/world/americas/brazil-alert-after-third-case-omicron-variant-2021-12-01/},
  year         = {2021},
 note = {Accessed on 01/09/2026}
}

@misc{primeiravacinaSP,
  author       = {Bianka Vieira},
  title        = {{'I Always Want to Do Good for My Race', Says The First Vaccinated against Covid in Brazil}},
  howpublished = {Folha de S.Paulo, https://folha.com/o8t1fdrg},
  year         = {2021},
 note = {Accessed on 01/09/2026}
}

@misc{primeiravacina,
  author       = {{Natalie Cancian, Renato Machado}},
  title        = {{Anvisa Approves Definitive Registration of Pfizer Vaccine against Covid}},
  howpublished = {Folha de S.Paulo, https://folha.com/e0cxpbns},
  year         = {2021},
 note = {Accessed on 01/09/2026}
}

@inproceedings{schneider-etal-2020-biobertpt,
    title = "{B}io{BERT}pt - A {P}ortuguese Neural Language Model for Clinical Named Entity Recognition",
    author = "Schneider, Elisa Terumi Rubel  and
      de Souza, Jo{\~a}o Vitor Andrioli  and
      Knafou, Julien  and
      Oliveira, Lucas Emanuel Silva e  and
      Copara, Jenny  and
      Gumiel, Yohan Bonescki  and
      Oliveira, Lucas Ferro Antunes de  and
      Paraiso, Emerson Cabrera  and
      Teodoro, Douglas  and
      Barra, Cl{\'a}udia Maria Cabral Moro",
    booktitle = "Proc. of the Clinical Natural Language Processing Workshop",
    month = nov,
    year = "2020",
    url = "https://www.aclweb.org/anthology/2020.clinicalnlp-1.7",
    pages = "65--72",
}

@article{roberta,
  title = "Unsupervised Cross-lingual Representation Learning at Scale",
    author = "Conneau, Alexis  and
      Khandelwal, Kartikay  and
      Goyal, Naman  and
      Chaudhary, Vishrav  and
      Wenzek, Guillaume  and
      Guzm{\'a}n, Francisco  and
      Grave, Edouard  and
      Ott, Myle  and
      Zettlemoyer, Luke  and
      Stoyanov, Veselin",
    editor = "Jurafsky, Dan  and
      Chai, Joyce  and
      Schluter, Natalie  and
      Tetreault, Joel",
    booktitle = "Proceedings of the 58th Annual Meeting of the Association for Computational Linguistics",
    month = jul,
    year = "2020",
    address = "Online",
    publisher = "Association for Computational Linguistics",
    url = "https://aclanthology.org/2020.acl-main.747/",
    doi = "10.18653/v1/2020.acl-main.747",
    pages = "8440--8451"
}

@inproceedings{souza2020bertimbau,
  author    = {F{\'a}bio Souza and
               Rodrigo Nogueira and
               Roberto Lotufo},
  title     = {{BERT}imbau: pretrained {BERT} models for {B}razilian {P}ortuguese},
  booktitle = {Brazilian Conference on Intelligent Systems},
  year      = {2020}
}

@article{Kowsari_2019,
   title={Text Classification Algorithms: A Survey},
   volume={10},
   ISSN={2078-2489},
   url={http://dx.doi.org/10.3390/info10040150},
   DOI={10.3390/info10040150},
   number={4},
   journal={Information},
   publisher={MDPI AG},
   author={Kowsari, Kamran and Jafari Meimandi, Kiana and Heidarysafa, Mojtaba and Mendu, Sanjana and Barnes, Laura and Brown, Donald},
   year={2019},
   month=apr, pages={150} }

@inproceedings{hossain-etal-2020-covidlies,
    title = "{COVIDL}ies: Detecting {COVID}-19 Misinformation on Social Media",
    author = "Hossain, Tamanna  and
      Logan IV, Robert L.  and
      Ugarte, Arjuna  and
      Matsubara, Yoshitomo  and
      Young, Sean  and
      Singh, Sameer",
    booktitle = "Proc. of the Workshop on {NLP} for {COVID}-19 at {EMNLP}",
    month = dec,
    year = "2020",
    url = "https://aclanthology.org/2020.nlpcovid19-2.11/",
    doi = "10.18653/v1/2020.nlpcovid19-2.11"
}

@inproceedings{chen2021mmcovar,
title = "Mmcovar: multimodal COVID-19 vaccine focused data repository for fake news detection and a baseline architecture for classification",
author = "Chen, Mingxuan and Chu, Xinqiao and Subbalakshmi, KP",
booktitle = "Proc. of the IEEE/ACM Int. Conference on Advances in Social Networks Analysis and Mining",
year = "2021",
pages = "31--38",}

@inproceedings{poddar2022caves,
  title={CAVES: A dataset to facilitate Explainable Classification and Summarization of Concerns towards COVID Vaccines},
  author={Poddar, Soham and Samad, Azlaan Mustafa and Mukherjee, Rajdeep and Ganguly, Niloy and Ghosh, Saptarshi},
  booktitle={Proc. of the Int. ACM SIGIR Conference on Research and Development in Information Retrieval},
  year={2022}
}

@article{Mu_Jin_Grimshaw_Scarton_Bontcheva_Song_2023, title={VaxxHesitancy: A Dataset for Studying Hesitancy towards COVID-19 Vaccination on Twitter}, volume={17}, url={https://ojs.aaai.org/index.php/ICWSM/article/view/22213}, DOI={10.1609/icwsm.v17i1.22213}, abstractNote={Vaccine hesitancy has been a common concern, probably since vaccines were created and, with the popularisation of social media, people started to express their concerns about vaccines online alongside those posting pro- and anti-vaccine content. Predictably, since the first mentions of a COVID-19 vaccine, social media users posted about their fears and concerns or about their support and belief into the effectiveness of these rapidly developing vaccines. Identifying and understanding the reasons behind public hesitancy towards COVID-19 vaccines is important for policy markers that need to develop actions to better inform the population with the aim of increasing vaccine take-up. In the case of COVID-19, where the fast development of the vaccines was mirrored closely by growth in anti-vaxx disinformation, automatic means of detecting citizen attitudes towards vaccination became necessary. This is an important computational social sciences task that requires data analysis in order to gain in-depth understanding of the phenomena at hand. Annotated data is also necessary for training data-driven models for more nuanced analysis of attitudes towards vaccination. To this end, we created a new collection of over 3,101 tweets annotated with users’ attitudes towards COVID-19 vaccination (stance). Besides, we also develop a domain-specific language model (VaxxBERT) that achieves the best predictive performance (73.0 accuracy and 69.3 F1-score) as compared to a robust set of baselines. To the best of our knowledge, these are the first dataset and model that model vaccine hesitancy as a category distinct from pro- and anti-vaccine stance.}, number={1}, journal={Proc. of the Int. AAAI Conference on Web and Social Media}, author={Mu, Yida and Jin, Mali and Grimshaw, Charlie and Scarton, Carolina and Bontcheva, Kalina and Song, Xingyi}, year={2023}, month={Jun.}, pages={1052-1062} }

@article{Crupi_Mejova_Tizzani_Paolotti_Panisson_2022, title={Echoes through Time: Evolution of the Italian COVID-19 Vaccination Debate}, volume={16}, url={https://ojs.aaai.org/index.php/ICWSM/article/view/19276}, DOI={10.1609/icwsm.v16i1.19276}, abstractNote={Italy was the first European country to be hit by COVID-19 in the early 2020, since then losing over 100,000 people to the disease. By the end of the vaccination campaign of 2021, 81% of the public received at least one dose. These dramatic developments were accompanied by a rigorous discussion around vaccination, both about its urgency and its possible negative effects. Twitter is one of the most popular social media platforms in the country, but pre-pandemic vaccination debate has been shown to be polarized and siloed into echo chambers. It is thus imperative to understand the nature of this discourse, with a specific focus on the vaccination hesitant individuals, whose healthcare decisions may affect their communities and the country at large. In this study we ask, how has the Italian discussion around vaccination changed during the COVID-19 pandemic, and have the unprecedented events of 2020-2021 been able to break the echo chamber around this topic? We use a Twitter dataset spanning September 2019 - November 2021 to examine the state of polarization around vaccination. We propose a hierarchical clustering approach to find the largest communities in the endorsement networks of different time periods, and manually illustrate that it produces communities of users sharing a stance. Examining the structure of these networks, as well as textual content of their interactions, we find the stark division between supporters and hesitant individuals to continue throughout the vaccination campaign. However, we find an increasing commonality in the topical focus of the vaccine supporters and vaccine hesitant, pointing to a possible common set of facts the two sides may agree on. Still, we discover a series of concerns voiced by the hesitant community, ranging from unfounded conspiracies (microchips in vaccines) to public health policy discussion (vaccine passport limitations). We recommend an ongoing surveillance of this debate, especially to uncover concerns around vaccination before the public health decisions and official messaging are made public.}, number={1}, journal={Proc. of the Int. AAAI Conference on Web and Social Media}, author={Crupi, Giuseppe and Mejova, Yelena and Tizzani, Michele and Paolotti, Daniela and Panisson, André}, year={2022}, month={May}, pages={102-113} }

@article{Hayawi2022ANTiVax,
  author    = {Hayawi, K. and Shahriar, S. and Serhani, M. A. and Taleb, I. and Mathew, S. S.},
  title     = {{ANTi-Vax: a novel Twitter dataset for COVID-19 vaccine misinformation detection}},
  journal   = {Public Health},
  volume    = {203},
  pages     = {23--30},
  year      = {2022},
  month     = feb,
  doi       = {10.1016/j.puhe.2021.11.022},
  pmid      = {35016072},
  pmcid     = {PMC8648668},
  note      = {Epub 2021 Dec 7}
}

@inproceedings{zarei2023vaxculturedatasetstudyingvaccine,
  author={Zarei, Mohammad Reza and Christensen, Michael and Everts, Sarah and Komeili, Majid},
  booktitle={Int. Joint Conference on Neural Networks}, 
  title={Vax-Culture: A Dataset for Studying Vaccine Discourse on Twitter}, 
  year={2023},
  volume={},
  number={},
  pages={1-8},
  doi={10.1109/IJCNN54540.2023.10191981}}

@article{Poddar_Mondal_Misra_Ganguly_Ghosh_2022, title={Winds of Change: Impact of COVID-19 on Vaccine-Related Opinions of Twitter Users}, volume={16}, url={https://ojs.aaai.org/index.php/ICWSM/article/view/19334}, DOI={10.1609/icwsm.v16i1.19334}, abstractNote={Today, administering COVID-19 vaccines at a societal scale has been deemed as the most appropriate way to defend against the COVID-19 pandemic. This global vaccination drive naturally fueled a possibility of pro-vax and anti-vax users strongly expressing their supports and concerns regarding the vaccines in online social media platforms. Understanding this online discourse is crucial for policy makers. This understanding is likely to impact the success of vaccination drives and might even impact the final outcome of our fight against the pandemic. The goal of this work is to improve this understanding using the lens of Twitter-discourse data. We first develop a classifier to categorize users according to their vaccine-related stance with high precision (97%). Using this method we detect and investigate specific user-groups who posted about vaccines in pre-COVID and COVID times. We specifically identify distinct topics that these users talk about, and investigate how vaccine-related discourse has changed between pre-COVID times and COVID times. Finally, for the first time, we investigate the change of vaccine-related stances in Twitter users and shed light on potential reasons for such changes in stance.}, number={1}, journal={Proc. of the Int. AAAI Conference on Web and Social Media}, author={Poddar, Soham and Mondal, Mainack and Misra, Janardan and Ganguly, Niloy and Ghosh, Saptarshi}, year={2022}, month={May}, pages={782-793} }

@article{Giovanni_Pierri_Torres-Lugo_Brambilla_2022, title={VaccinEU: COVID-19 Vaccine Conversations on Twitter in French, German and Italian}, volume={16}, url={https://ojs.aaai.org/index.php/ICWSM/article/view/19374}, DOI={10.1609/icwsm.v16i1.19374}, abstractNote={Despite the increasing limitations for unvaccinated people, in many European countries there is still a non-negligible fraction of individuals who refuse to get vaccinated against SARS-CoV-2, undermining governmental efforts to eradicate the virus. We study the role of online social media in influencing individuals’ opinion towards getting vaccinated by designing a large-scale collection of Twitter messages in three different languages -- French, German and Italian -- and providing public access to the data collected. Focusing on the European context, our VaccinEU dataset aims to help researchers to better understand the impact of online (mis)information about vaccines and design more accurate communication strategies to maximize vaccination coverage. Data can be fully accessed in a Dataverse repository and a GitHub repository.}, number={1}, journal={Proc. of the Int. AAAI Conference on Web and Social Media}, author={Giovanni, Marco Di and Pierri, Francesco and Torres-Lugo, Christopher and Brambilla, Marco}, year={2022}, month={May}, pages={1236-1244} }

@inproceedings{webmedia_disinfo_review,
author = {João Olivetti and Ana Bomfim and Marcos Oliveira and Helom Marques and Juan Avelar and Julio Reis and Philipe Melo},
 title = { O Fenômeno da Desinformação no WhatsApp, Telegram e Outras Plataformas de Mensagens Instantâneas: Uma Revisão Sistemática da Literatura (in Portuguese)},
 booktitle = {Proc. of the Brazilian Symposium on Multimedia and the Web},
 year = {2025},
 pages = {661--675},
 doi = {10.5753/webmedia.2025.16058}
}

@Article{covid_misinfo_1,
author="Skafle, Ingjerd
and Nordahl-Hansen, Anders
and Quintana, Daniel S
and Wynn, Rolf
and Gabarron, Elia",
title="Misinformation About COVID-19 Vaccines on Social Media: Rapid Review",
journal="J Med Internet Res",
year="2022",
month="Aug",
day="4",
volume="24",
number="8",
pages="e37367",
issn="1438-8871",
doi="10.2196/37367",
url="https://www.jmir.org/2022/8/e37367",
url="https://doi.org/10.2196/37367",
url="http://www.ncbi.nlm.nih.gov/pubmed/35816685"
}

@Article{covid_misinfo_2,
title = {The impact of misinformation on the COVID-19 pandemic},
journal = {AIMS Public Health},
volume = {9},
number = {2},
pages = {262-277},
year = {2022},
issn = {2327-8994},
doi = {10.3934/publichealth.2022018},
url = {https://www.aimspress.com/article/doi/10.3934/publichealth.2022018},
author = {Maria Mercedes Ferreira Caceres and Juan Pablo Sosa and Jannel A Lawrence and Cristina Sestacovschi and Atiyah Tidd-Johnson and Muhammad Haseeb UI Rasool and Vinay Kumar Gadamidi and Saleha Ozair and Krunal Pandav and Claudia Cuevas-Lou and Matthew Parrish and Ivan Rodriguez and Javier Perez Fernandez},
}

@Article{antivax_desinfo_1,
author={Loomba, Sahil
and de Figueiredo, Alexandre
and Piatek, Simon J.
and de Graaf, Kristen
and Larson, Heidi J.},
title={Measuring the impact of COVID-19 vaccine misinformation on vaccination intent in the UK and USA},
journal={Nature Human Behaviour},
year={2021},
month={Mar},
day={01},
volume={5},
number={3},
pages={337-348},
issn={2397-3374},
doi={10.1038/s41562-021-01056-1},
url={https://doi.org/10.1038/s41562-021-01056-1}
}

@Article{antivax_desinfo_2,
author="Zhao, Sihong
and Hu, Simeng
and Zhou, Xiaoyu
and Song, Suhang
and Wang, Qian
and Zheng, Hongqiu
and Zhang, Ying
and Hou, Zhiyuan",
title="The Prevalence, Features, Influencing Factors, and Solutions for COVID-19 Vaccine Misinformation: Systematic Review",
journal="JMIR Public Health Surveill",
year="2023",
month="Jan",
day="11",
volume="9",
pages="e40201",
issn="2369-2960",
doi="10.2196/40201",
url="https://publichealth.jmir.org/2023/1/e40201",
url="https://doi.org/10.2196/40201",
url="http://www.ncbi.nlm.nih.gov/pubmed/36469911"
}

@inproceedings{Resende-WWW2019,
author = {Resende, Gustavo and Melo, Philipe and Sousa, Hugo and Messias, Johnnatan and Vasconcelos, Marisa and Almeida, Jussara and Benevenuto, Fabr\'{\i}cio},
title = {(Mis)Information Dissemination in WhatsApp: Gathering, Analyzing and Countermeasures},
year = {2019},
booktitle = {The World Wide Web Conference},
pages = {818–828}
}

@article{melo2019whatsapp, 
    title={WhatsApp Monitor: A Fact-Checking System for WhatsApp},
    author={Melo, Philipe and Messias, Johnnatan and Resende, Gustavo and Garimella, Kiran and Almeida, Jussara and Benevenuto, Fabrício}, 
    journal={Proc. of the Int'l AAAI Conference on Web and Social Media}, 
    volume={13}, 
    url={https://ojs.aaai.org/index.php/ICWSM/article/view/3271}, 
    DOI={10.1609/icwsm.v13i01.3271}, 
    number={01}, 
    year={2019}, 
    month={Jul.}, 
    pages={676-677} 
}

@inproceedings{melo2019complexnet,
  title={Can WhatsApp Counter Misinformation by Limiting Message Forwarding?},
  author={Melo, Philipe and Vieira, Carolina Coimbra and Garimella, Kiran and de Melo, Pedro OS Vaz and Benevenuto, Fabr{\'\i}cio},
  booktitle={Int'l Conference on Complex Networks and Their Applications},
  pages={372--384},
  year={2019}
}

@article{benevenuto2024misinformation,
  title={Misinformation Campaigns through WhatsApp and Telegram in Presidential Elections in Brazil},
  author={Benevenuto, Fabr{\'\i}cio and Melo, Philipe},
  journal={Comm. of the ACM},
  volume={67},
  number={8},
  pages={72--77},
  year={2024},
  publisher={ACM New York, NY, USA}
}

@article{reis2020can,
  title={Can WhatsApp benefit from debunked fact-checked stories to reduce misinformation?},
  author={Reis, Julio CS and Melo, Philipe and Garimella, Kiran and Benevenuto, Fabr{\'\i}cio},
  journal={Harvard Kennedy School Misinformation Review},
  year={2020}
}

@inproceedings{reimers-2019-sentence-bert,
  title = "Sentence-BERT: Sentence Embeddings using Siamese BERT-Networks",
  author = "Reimers, Nils and Gurevych, Iryna",
  booktitle = "Proc. of the Conference on Empirical Methods in Natural Language Processing",
  month = "11",
  year = "2019"
}

@article{douze2024faiss,
  title={The faiss library},
  author={Douze, Matthijs and Guzhva, Alexandr and Deng, Chengqi and Johnson, Jeff and Szilvasy, Gergely and Mazar{\'e}, Pierre-Emmanuel and Lomeli, Maria and Hosseini, Lucas and J{\'e}gou, Herv{\'e}},
  journal={IEEE Transactions on Big Data},
  year={2025}
}

@article{internet_behavior,
author = {Benjamin C. Storm and Sean M. Stone and Aaron S. Benjamin},
title = {Using the Internet to access information inflates future use of the Internet to access other information},
journal = {Memory},
volume = {25},
number = {6},
pages = {717--723},
year = {2017},
publisher = {Routledge},
doi = {10.1080/09658211.2016.1210171},
note ={PMID: 27424847},
URL = { https://doi.org/10.1080/09658211.2016.1210171 }
}

@article{elections1,
author = {Jacinta Mwende Maweu},
title = {“Fake Elections”? Cyber Propaganda, Disinformation and the 2017 General Elections in Kenya},
journal = {African Journalism Studies},
volume = {40},
number = {4},
pages = {62--76},
year = {2019},
publisher = {Routledge},
doi = {10.1080/23743670.2020.1719858},
URL = { 
        https://doi.org/10.1080/23743670.2020.1719858
}
}

@article{elections2, 
title={A Dataset of Fact-Checked Images Shared on WhatsApp During the Brazilian and Indian Elections}, 
volume={14}, url={https://ojs.aaai.org/index.php/ICWSM/article/view/7356}, DOI={10.1609/icwsm.v14i1.7356}, abstractNote={&lt;p&gt;Recently, messaging applications, such as WhatsApp, have been reportedly abused by misinformation campaigns, especially in Brazil and India. A notable form of abuse in WhatsApp relies on several manipulated images and memes containing all kinds of fake stories. In this work, we performed an extensive data collection from a large set of WhatsApp publicly accessible groups and fact-checking agency websites. This paper opens a novel dataset to the research community containing fact-checked fake images shared through WhatsApp for two distinct scenarios known for the spread of fake news on the platform: the 2018 Brazilian elections and the 2019 Indian elections.&lt;/p&gt;}, number={1}, journal={Proc. of the Int. AAAI Conference on Web and Social Media}, author={Reis, Julio C. S. and Melo, Philipe and Garimella, Kiran and Almeida, Jussara M. and Eckles, Dean and Benevenuto, Fabrício}, year={2020}, month={May}, pages={903-908} }

@inproceedings{reis2023helping,
  title={Helping fact-checkers identify fake news stories shared through images on whatsapp},
  author={Reis, Julio CS and Melo, Philipe and Bel{\'e}m, Fabiano and Murai, Fabricio and Almeida, Jussara M and Benevenuto, Fabricio},
  booktitle={Proc. of the Brazilian Symposium on Multimedia and the Web},
  pages={159--167},
  year={2023}
}

@article{elections3,
author = {Marlene Mauk and Max Grömping},
title ={Online Disinformation Predicts Inaccurate Beliefs About Election Fairness Among Both Winners and Losers},
journal = {Comparative Political Studies},
volume = {57},
number = {6},
pages = {965-998},
year = {2024},
doi = {10.1177/00104140231193008},

URL = { 
    
        https://doi.org/10.1177/00104140231193008
}
}

@ARTICLE{overload1,
  
AUTHOR={Huang, Qing  and Lei, Sihan  and Ni, Binbin },
         
TITLE={Perceived Information Overload and Unverified Information Sharing on WeChat Amid the COVID-19 Pandemic: A Moderated Mediation Model of Anxiety and Perceived Herd},
        
JOURNAL={Frontiers in Psychology},
        
VOLUME={Volume 13 - 2022},

YEAR={2022},

URL={https://www.frontiersin.org/journals/psychology/articles/10.3389/fpsyg.2022.837820},

DOI={10.3389/fpsyg.2022.837820},

ISSN={1664-1078}}

@article{overload2,
title = {Information overload and fake news sharing: A transactional stress perspective exploring the mitigating role of consumers’ resilience during COVID-19},
journal = {Journal of Retailing and Consumer Services},
volume = {61},
pages = {102555},
year = {2021},
issn = {0969-6989},
doi = {https://doi.org/10.1016/j.jretconser.2021.102555},
url = {https://www.sciencedirect.com/science/article/pii/S0969698921001211},
author = {Alena Bermes}
}

@Article{osn_misinfo1,
author="Suarez-Lledo, Victor
and Alvarez-Galvez, Javier",
title="Prevalence of Health Misinformation on Social Media: Systematic Review",
journal="J Med Internet Res",
year="2021",
month="Jan",
day="20",
volume="23",
number="1",
pages="e17187",
issn="1438-8871",
doi="10.2196/17187",
url="http://www.jmir.org/2021/1/e17187/",
url="https://doi.org/10.2196/17187",
url="http://www.ncbi.nlm.nih.gov/pubmed/33470931"
}

@article{osn_misinfo2,
title = {Misinformation on social platforms: A review and research Agenda},
journal = {Technology in Society},
volume = {78},
pages = {102654},
year = {2024},
issn = {0160-791X},
doi = {https://doi.org/10.1016/j.techsoc.2024.102654},
url = {https://www.sciencedirect.com/science/article/pii/S0160791X24002021},
author = {Neha Chaudhuri and Gaurav Gupta and Mehdi Bagherzadeh and Tugrul Daim and Haydar Yalcin}
}

@misc{UNICEF_WHO_2025,
  title        = {Global childhood vaccination coverage holds steady, yet over 14 million infants remain unvaccinated},
  author       = {{World Health Organization} and {UNICEF}},
  year         = {2025},
  howpublished = {https://www.who.int/news/item/15-07-2025-global-childhood-vaccination-coverage-holds-steady-yet-over-14-million-infants-remain-unvaccinated-who-unicef},
  note = {Accessed on 01/09/2026}
}

@inproceedings{sbbd_vaccine,
 author = {Antônio Martins and Lucas Cabral and Pedro Jorge Mourão and José Monteiro and Javam Machado},
 title = {Detection of Misinformation about COVID-19 in Brazilian Portuguese WhatsApp Messages Using Deep Learning},
 booktitle = {Proc. of the Brazilian Symposium on Databases},
 location = {Rio de Janeiro},
 year = {2021},
 issn = {2763-8979},
 pages = {85--96},
 doi = {10.5753/sbbd.2021.17868},
 url = {https://sol.sbc.org.br/index.php/sbbd/article/view/17868}
}
}

\end{document}